\documentclass{elsart}
\usepackage{epsfig}
\usepackage{amssymb}

\begin{document}
\begin{frontmatter}

\title{Central Limit behavior in the Kuramoto model at the 'Edge of Chaos'} 

\author
{Giovanna Miritello}  \ead{giovanna.miritello@ct.infn.it}
\address {Departamento de Matematicas, Universidad Carlos III de Madrid, Avenida de la Universidad 30, 28911 Legan\'es, Spain}

\author
{Alessandro Pluchino} \ead{alessandro.pluchino@ct.infn.it}
\address{Dipartimento di Fisica e Astronomia,  Universit\'a di Catania,\\
and INFN sezione di Catania,  Via S. Sofia 64, I-95123 Catania, Italy} 

\author
{Andrea Rapisarda} \ead{andrea.rapisarda@ct.infn.it}
\address{Dipartimento di Fisica e Astronomia,  Universit\'a di Catania,\\
and INFN sezione di Catania,  Via S. Sofia 64, I-95123 Catania, Italy} 

%
%

\begin{abstract}


We study the relationship between chaotic behavior and the Central Limit Theorem (CLT) 
in the Kuramoto model. We calculate sums of angles at equidistant times along  deterministic 
trajectories of single oscillators and we show that, when chaos is sufficiently strong , the Pdfs of the sums  
tend to a Gaussian, consistently with the standard CLT. On the other hand, when the system is 
at the "edge of chaos" (i.e. in a regime with vanishing Lyapunov exponents), robust $q$-Gaussian-like attractors 
naturally emerge, consistently with recently proved generalizations of the CLT.

\end{abstract}



\end{frontmatter}

\section{Introduction}

The relationship between weak chaos and the Central Limit Theorem (CLT) 
has been recently explored by several authors, both in low-dimensional chaotic maps 
\cite{CLT-maps1,CLT-maps2,CLT-maps3,CLT-maps4} and in long-range many-body 
systems \cite{CLT-hmf1,CLT-hmf2}. 
The Standard Central Limit Theorem (CLT) states that the (rescaled)
sum of $n$ independent random variables with finite variance has
a Gaussian distribution in the limit $n\rightarrow\infty$. On the other hand, 
the theorem does not hold when the variables under consideration are
weakly mixing and strongly correlated, as it happens in systems with zero or 
vanishing Lyapunov exponents, i.e. systems at the "edge of chaos". 
In order to consider  this case,  it has recently been introduced a 
q-generalization of the standard CLT, which states that for certain classes of strongly
correlated random variables the (rescaled) sum approaches a $q$-Gaussian
distribution \cite{q-CLT}. The latter is a generalization of a Gaussian
distribution which emerges in the context of the nonextensive statistical
mechanics \cite{tsallis0}. It is defined as $G_{q}(\beta,x)=A(q,\beta) [1-(1-q) \beta x^{2}]^{1/(1-q)}$, 
where $q$ is the entropic  index (such that for $q=1$ one recovers
the usual Gaussian), $\beta$ a parameter related to the width
of the distribution and $A(q,\beta)$ is a normalizing constant.
\\
In previous works \cite{CLT-hmf1,CLT-hmf2} we showed that $q$-Gaussian 
attractors exist for the rescaled sums of the angular velocities, calculated 
along deterministic trajectories of single oscillators (time-average Pdfs), in the 
weakly chaotic quasi-stationary (QSS) regime of the Hamiltonian Mean Field (HMF) model
\cite{hmf-model}.
On the other hand, we also pointed out \cite{kura-hmf1,kura-hmf2} several
analogies (metastability, continuous phase transition, chaotic behavior) 
between the HMF model and the Kuramoto model \cite{kuramoto},
probably due to their common origin from a more general model of coupled
driven-damped pendula.
In this paper we present new numerical results which extend these
analogies to the CLT features. In particular we will show that also in
the context of the Kuramoto model a very weakly chaotic microscopic dynamics
with vanishing Lyapunov exponents gives rise to the emergence of robust $q$-Gaussian-like 
attractors, which  disappear when the level of chaos increases, thus restoring the standard 
CLT behavior.
\\
In the first part of the paper we describe the Kuramoto model phase transition
scenario, with specific attention to the chaotic behavior and to the role 
played by the distribution of the natural frequencies  (uniform or Gaussian).
In the second part we discuss the numerical results on the CLT behavior,
which are strongly influenced by the interplay between chaos and synchronization.

\section{Phase transitions, Chaos and Metastability in the Kuramoto model}

The Kuramoto model \cite{kuramoto,strogatz,acebron} describes the following dissipative dynamics 
of a system of $N$ sinusoidally coupled oscillators:
\begin{equation}
\dot{\theta}_{i}=\omega_{i}+\frac{K}{N}\sum_{j=1}^{N}\sin(\theta_{j}-\theta_{i}),
\label{eq-moto-kura}
\end{equation}
where $\theta_{i}$ and $\omega_{i}$ are, respectively, the angle (phase) and the natural frequency 
of the $i$-th oscillator, with $i=1,...,N$, while $K\geq0$ is the coupling strenght. 
The natural frequencies are time-independent and randomly chosen from a symmetric, 
unimodal distribution $g(\omega)$. They represent a driving term
that, considering  the oscillators as particles moving on the unit circle, forces 
each particle to rotate at its own frequency regardless of the other particles. 
On the other hand, the sinusoidal term, tuned by the coupling constant $K$, tends to
synchronize the particles, so that, for values of $K$ lower than a critical threshold $K_C$, 
the oscillators runs independently and incoherently, while for $K> K_C$ the system undergoes
a spontaneous phase transition towards a partially or fully synchronized state (depending on the value of $K$). 
Kuramoto \cite{kuramoto} showed analitically that, for a given unimodal distribution $g(\omega)$, 
the critical value of the coupling follows the expression $K_{C}=\frac{2}{\pi g(0)}$, being $g(0)$ the central value of the 
distribution. 
The degree of synchronization of the system is measured by the modulus $0 \le r \le  1$ of the dynamical 
vector order parameter $\mathbf{r}=re^{i\phi}=\frac{1}{N}\sum_{j=1}^{N}e^{i\theta_{j}}$,
where $\phi$ represents the centroid of the angles of the oscillators. 
\\
Using the order parameter $r$, eq. (\ref{eq-moto-kura}) can be rewritten in its mean field version as
$\dot{\theta}_{i}=\omega_{i}+Kr\sin(\phi-\theta_{i}),\:i=1,\,...,N$,
which allows to appreciate the role of both the natural frequencies $\omega_{i}$
and the absolute value of the product $Kr$ on determining the synchronization degree of the 
Kuramoto stationary states (i.e. those with $\dot{\theta}_{i}=0$).
In fact, as long as  $K$ is smaller than $K_C$, one has $\left|\omega_{i}\right|\geq Kr$ for all $i$ and  the oscillators drift at their own frequencies ($drifting$ oscillators) along the unit circle, so that 
 the system is incoherent and $r\sim0$.
When  $K$ increases above $K_C$, the system starts to synchronize and a cluster of $locked$ oscillators, 
those with $\left|\omega_{i}\right|\leq Kr$, appears. In correspondence, $r$ starts to grow
(indicating partial synchronization) and the number of drifting oscillators decreases more and more until, 
for large value of $K$ and for $r>0.8$, the system becomes fully synchronized and all the oscillators
rotate at the average frequency $\Omega$, a quantity which is conserved  during the dynamical evolution.

\begin{figure}
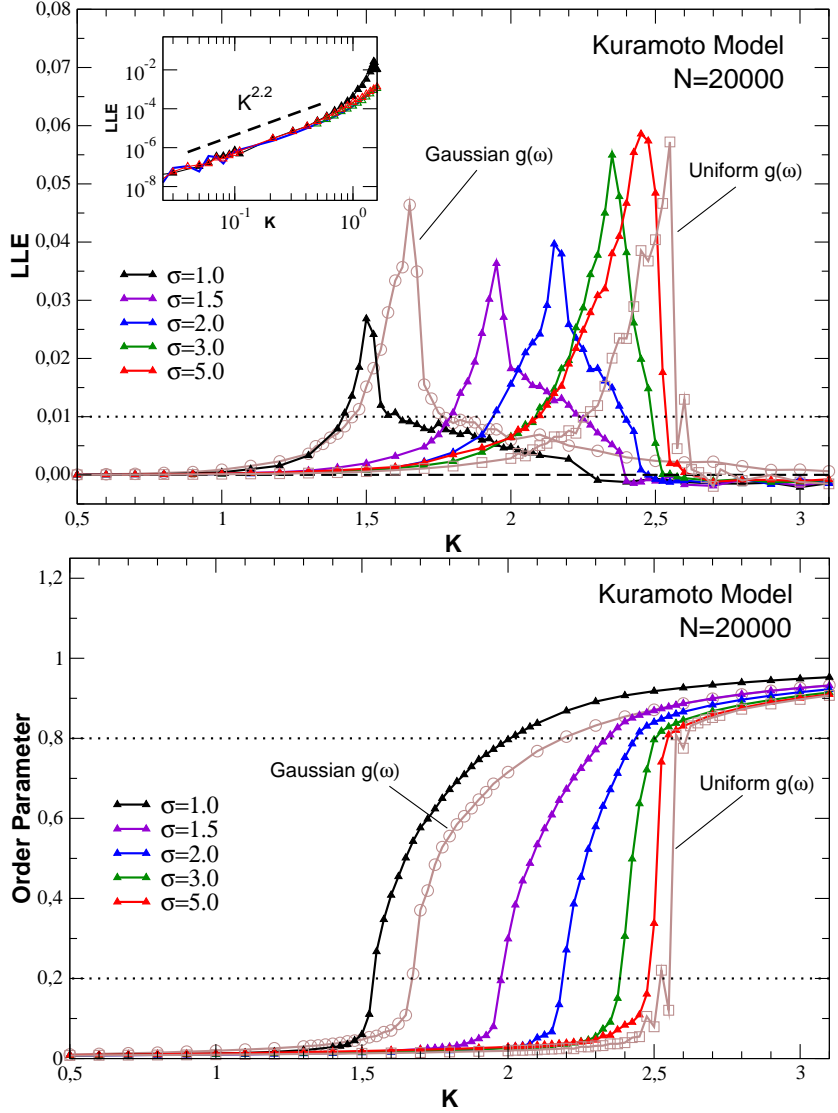
  
\begin{center}
\epsfig{figure=fig1a.eps,width=11truecm,angle=0}
\epsfig{figure=fig1b.eps,width=11truecm,angle=0}
\end{center}
\caption{(Lower panel) The asymptotic order parameter $r$ of the Kuramoto model is plotted as a function of the coupling $K$ for a system of $N=20000$ oscillators and for several bounded Gaussian distributions $g(\omega)$, with different standard deviations $\sigma$ and $\omega\in[-2,2]$.   
Increasing $\sigma$ from $1$ to $5$ (triangles lines, from left to right), the $g(\omega)$'s changes continuously from a Gaussian-like to a uniform-like distribution and, in correspondence, the phase transition changes continuously from second-order-like to first-order-like, while the critical value of the coupling $K_C$ shift to the left. 
The curves corresponding to a true Gaussian distribution (open circles) and a true uniform one (open squares) are also reported for comparison. 
(Upper panel) The Largest Lyapunov Exponent is plotted as function of $K$ for the same cases showed in the upper panel. Increasing $\sigma$ its behavior changes continuously (from left to right) from the characteristic, slowly decreasing, peak of a Gaussian $g(\omega)$ to the abruptly vanishing peak characteristic of the uniform one. Each dot represents an average over $10$ runs.
In the inset we  show the power-law scaling of the $LLE$ for small values of $K$. The  $LLE$ goes to zero as $7\cdot 10^5 \cdot K^{2.2 } $.
See text. } 
\label{fig1}
\end{figure}

In ref.\cite{kura-hmf2}, confirming and extending previous studies of other authors \cite{pazo},
we showed that the distribution $g(\omega)$ plays a fundamental
role for fixing the features of the phase transition in the Kuramoto model, even for large system sizes.
In particular we showed that the shape of the phase transition can be slowly changed from 
a continuous 2nd-order-like one (very similar to the HMF phase transition and
corresponding to a Gaussian $g(\omega)$), to an abrupt 1st-order-like one 
(with coexistence of phases and corresponding to a uniform $g(\omega)$), by tuning
the standard deviation of a bounded Gaussian distribution of the natural frequencies. 
\\
We report an example of such a behavior in Fig.1, where the phase transitions for a system of 
$N=20000$ are plotted for several values of the standard deviation $\sigma$ of 
bounded Gaussian distributions $g(\omega)$'s, with $\omega\in[-2,2]$ (examples of 
these distributions can be found in the insets of Fig.2).  
In the lower panel, the asymptotic order parameter $r$ is plotted as function of $K$ 
for five increasing values of $\sigma$ (namely $\sigma=1,1.5,2,3,5$, see triangles lines). 
The limiting cases $\sigma=1$ and $\sigma=5$ approximate, respectively, the true Gaussian 
and the true uniform $g(\omega)$ distributions, which are also reported for comparison 
(lines with open circles and open squares, respectively).
\\
In the upper panel, the behavior of the corresponding largest Lyapunov exponent ($LLE$) is 
reported for all these distributions.
As we already showed in ref.\cite{kura-hmf2}, the partially synchronized regime of the Kuramoto model, 
characterized by 
the simultaneous presence of drifting and locked oscillators and by an order parameter 
$0<r<0.8$, exhibits a positive $LLE$ with a peak just in coincidence of the phase transition, 
indicating  the presence of strong chaos (see also \cite{maistrenko}).  
On the other hand, in the incoherent phase, when all the oscillators are drifting, we observe a vanishing (but positive) $LLE$, indicating a regime at the "edge of chaos". 
The inset of the upper panel of Fig.1 reveals also an asymptotic power-law scaling of the $LLE$, which 
 goes to zero as $K^{2.2}$, a further analogy with a similar behavior
observed in the HMF model \cite{PRL98}, although in that case the power-law exponent was different.
For the discussion of the next section it is important to note that, in the partially synchronized regime, 
the microscopic chaotic dynamics of the drifting oscillators enters in competition with the synchronization  term represented by the locked oscillators. 
The drifting term is prevalent provided that $r<0.2$: we call this region of the partially synchronized 
regime the  "fully chaotic" one, since it is also roughly characterized by a $LLE > 0.01$. On the contrary, for $r > 0.2$ the synchronization  term becomes relevant and, even if the system is still sligthly chaotic (with a $0 < LLE < 0.01$), the cluster of locked oscillators starts to dominate the dynamics. 
Finally, for $r > 0.8$, we are in the fully synchronized phase and all the oscillators
belong to the locked cluster which rotates  at the average constant frequency $\Omega$ 
(in both the panels of Fig.1, we report horizontal dotted lines in order  to mark  these different regions). 
Notice that in the fully synchronized regime the system is integrable and the $LLE$ vanishes again as in the incoherent phase, but in this case it can also be slightly negative. Furthermore, being $g(\omega)$ symmetric with $\omega\in[-2,2]$, one has in this case $\Omega\sim0$, therefore the locked cluster tends asymptotically to stop (we remind in this respect that the Kuramoto system is dissipative).

\begin{figure}  
\begin{center}
\epsfig{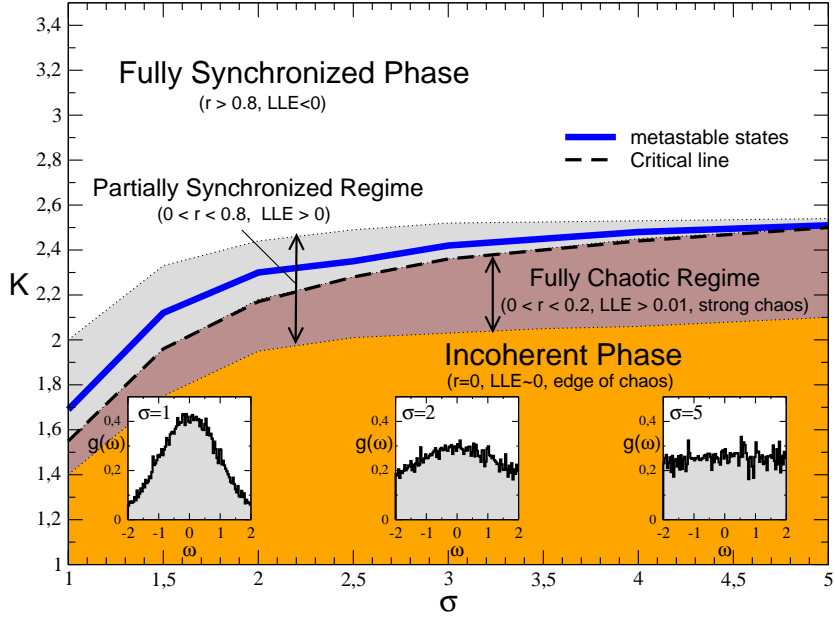}
\end{center}
\caption{Phase Diagram $K$ versus $\sigma$ for the Kuramoto model with $N=20000$. The critical line separates the fully synchronized phase (characterized by $r > 0.8$ and a slightly negative $LLE$) from the "edge of chaos" incoherent one (characterized by $r=0$ and a vanishing, but positive, $LLE$). The partially synchronized regime (characterized by $0 < r < 0.8$ and a positive $LLE$) is also visible between the two, around the phase transition. In particular, just below the critical line, we recognize a weakly synchronized sub-region of the partially synchronized regime, with values $0 < r < 0.2$ and $LLE>0.01$, that we call "fully chaotic regime" (see text).
In the insets, the bounded $g(\omega)$ distributions (with $\omega\in[-2,2]$) used in the simulations are plotted for three increasing values of $\sigma$. 
See text.  } 
\label{fig2}
\end{figure}

A schematic picture of the previous scenario is plotted in Fig.2, where the diagram $K$ versus $\sigma$
is reported together with the labels of the various phases and the corresponding values of $r$ and $LLE$
(this picture extends the diagram presented in ref.\cite{kura-hmf2}).
For $\sigma=1$ and $\sigma=5$ we have, respectively, the two estreme  cases of Gaussian-like and uniform-like
$g(\omega)$ (see the insets). 
The dashed critical line, drawn using the $LLE$ peaks in Fig.1, corresponds to a 2nd-order-like transition which - strictly speaking - would tend to a 1st-order-like one only in the limit $\sigma=\infty$, even if, as shown in Fig.1, the features of an abrupt 1st-order transition can be clearly recognized already for $\sigma=5$. 
The partially synchronized regime is visible between the incoherent and the fully synchronized 
phases, together with its sub-region corresponding to the fully chaotic regime, which is recognizable immediately below the critical line. Finally, just above the critical line, a thick full line (blue in color) going across
the partially synchronized regime indicates the presence of metastable states, see ref.\cite{kura-hmf1}.
In such states the interplay between locked and drifting oscillators hinders synchronization and does not allow
$r$ to reach immediately its maximum (stationary) value indicated by the curves in the lower panel of Fig.1. The situation is again similar to the one observed below the transition in the quasi-stationary regime of the HMF model, where some particles are locked in a single cluster while the others move inchoerently on the unit circle (see ref.\cite{kura-hmf2}).
\\  
In the next section we will expore the CLT behavior in the two limiting cases of a uniform and a Gaussian $g(\omega)$ distribution, where true 1st-order and 2nd-order phase transitions occur.
As shown in Fig.1, these two limiting situations are quite well approximated by the $\sigma=5$ and $\sigma=1$
bounded Gaussians cases, but do not exactly coincide with them. In particular, for the uniform and Gaussian 
cases, the value of $K_C$ is more consistent with the Kuramoto theoretical prediction, which becomes here, respectively, $K_{C}=\frac{8}{\pi}\sim2.54$ and $K_{C}=\frac{2\sqrt{2\pi}}{\pi}\sim1.59$, and both the $r(K)$ and $LLE(K)$ curves (open squares and open cirles lines) are slightly shifted on the right with respect to the $\sigma=5$ and $\sigma=1$ ones. In the following we will refer to these curves to analyze the CLT features of the Kuramoto model in its three peculiar regimes: incoherent, fully chaotic and fully synchronized.

\section{CLT behavior in the Kuramoto model}

Stimulating by the analogies between the Kuramoto and the HMF models found in refs.\cite{kura-hmf1,kura-hmf2},
and also by the observation of anomalous CLT behavior in the HMF context \cite{CLT-hmf1,CLT-hmf2},
we applied to the Kuramoto oscillators the procedure originally introduced in ref.\cite{CLT-maps1} for 
constructing probability density functions (CLT Pdfs) of quantities $y$'s expressed as a finite sum of stochastic variables, in order to identify possible violations of the standard CLT predictions. 
At variance with the HMF context, where we used angular velocities as stochastic variables, in this case the natural choice was to consider the angle variables of the N Kuramoto oscillators. Therefore we construct the rescaled sums $y$'s by picking out, for each oscillator, $n$ values of the angle $\theta_{i}$ at fixed intervals of time $\delta$ along the deterministic time evolution imposed by Eq.(\ref{eq-moto-kura}), i.e.:
\begin{equation}
y_i=\frac{1}{\sqrt{n}}\sum_{k=1}^{n}\theta_{i}(k\delta)\ \ \ \ \ with\ \ \ i=1,2,...,N.
\label{y-kura}
\end{equation}
In such a way, the product $\delta \times n$ will give the total simulation time over which the sum is calculated.
As explained in the introduction, for the standard CLT, we would expect a Gaussian $y$'s Pdf, but 
we already anticipated that the weakly chaotic microscopic dynamics, characterizing the Kuramoto
model with vanishing $LLE$, will give rise to the emergence of $q$-Gaussian-like attractors, which 
disappear when the level of chaos increases.
\\
First we show the results obtained for a uniform $g(\omega)$ distribution of
the natural frequencies, then we will pass to consider a Gaussian one.
In all the CLT numerical simulations we always consider 
a system of $N=20000$ oscillators, the same size used in Figs.1 and 2, which 
offers a good compromise between statistical effectiveness and computational demand.
We normalize also the $y's$ to zero mean and unitary variance variables, for allowing 
a better comparison with standard Gaussian and $q$-Gaussian curves.

\begin{figure}  
\begin{center}
\epsfig{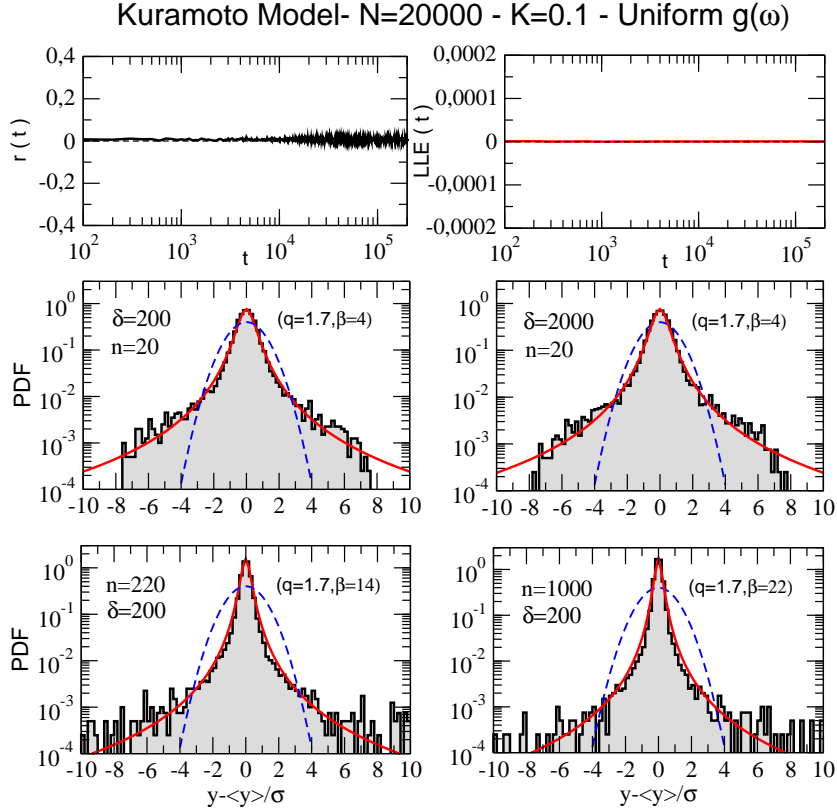}
\end{center}
\caption{Uniform $g(\omega)$: CLT Pdfs for a Kuramoto system with $N=20000$ and $K=0.1$ (in this case $K_C=2.54$), 
are reported for increasing values of $\delta$ at fixed $n$ (middle row) and for increasing 
values of $n$ at fixed $\delta$ (lower row): 
in both the cases robust fat tailed attractors appear, well fitted by $q$-Gaussians (full lines) 
with $q=1.7$ and different values of $\beta$. 
A standard Gaussian with unitary variance (dashed line) is reported for comparison in each plot. 
In the upper row of this and of the other figures by now on, the order parameter (on the left) 
and the LLE (on the right) are plotted for comparison as function of time, after a transient of 100 time-steps. See text. } 
\label{fig3}
\end{figure}

\subsection{Uniform $g(\omega)$ distribution}

Let us start with the case of a uniform $g(\omega)$ distribution
of the natural frequencies, with $\omega\in[-2,2]$. In the following we present 
several numerical simulations, one for each region of interest in agreement with the 
phase diagram of Fig.2 (see as an example the vertical section at $\sigma=5$, corresponding 
to the uniform-like $g(\omega)$ distribution case, quite similar to the true uniform case).
\\
In Fig.3 we show the $y$'s Pdfs of the same run at $K=0.1$, i.e. in the "edge of chaos" incoherent phase. We plot the Pdfs for increasing values of $\delta$ (200 and 2000) at fixed $n$=20 
(middle row) and for increasing values of $n$ (220 and 1000) at fixed $\delta$=200 (lower row).
The time evolution of both the order parameter $r$ and the $LLE$ is plotted in the upper row, after a transient of 100 time-steps. As  expected, both of them stay around zero for the entire time evolution (in particular, $LLE \sim 10^{-7}$). 
Correspondingly, one observes the occurrence of robust fat tailed Pdfs that persist increasing $\delta$  or $n$. The latter can be fitted very well by $q$-Gaussian curves (full lines) with $q=1.7$ and 
different values of $\beta$ reported in the figure. A standard Gaussian with unitary variance (dashed line) is also  plotted for comparison.

\begin{figure}  
\begin{center}
\epsfig{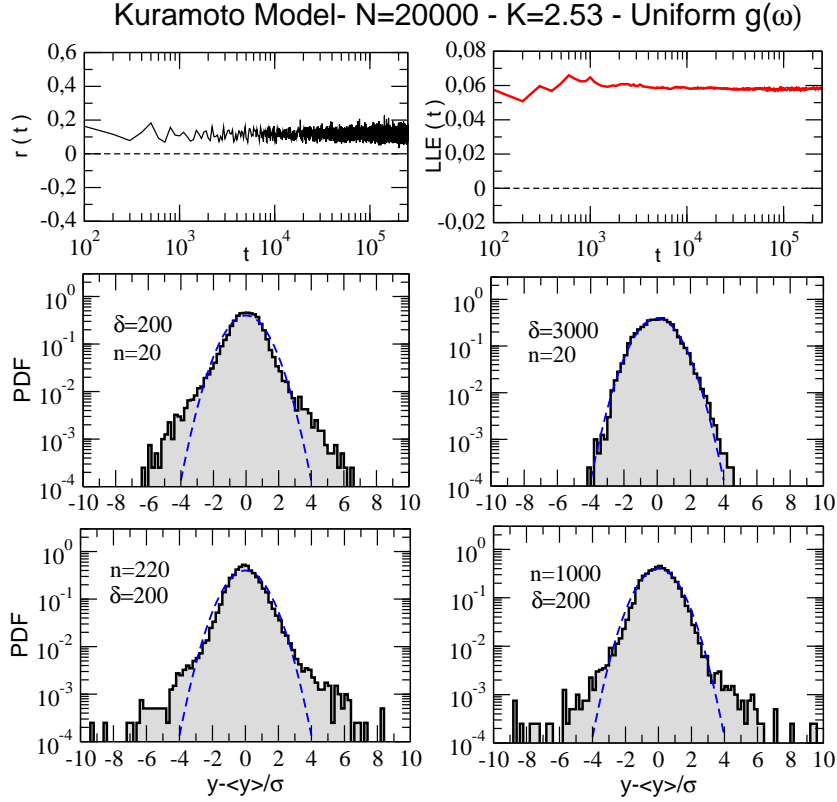}
\end{center}
\caption{Uniform $g(\omega)$, $N=20000$ and $K=2.53$ (in this case $K_C=2.54$). The fully chaotic regime tends produce 
Gaussian-like Pdfs, more visible increasing either $\delta$ or $n$. Standard Gaussian curves are
also reported for comparison. See text. } 
\label{fig4}
\end{figure}

Increasing $K$ up to  $K=2.53$ (very near to the critical value $K_C=2.54$ for the uniform $g(\omega)$ distribution and corresponding to the $LLE$ peak in the upper panel of Fig.1 - open squares line),  one enters into the fully chaotic regime: 
here the system is still quite incoherent ($0<r<0.2$) but the high level of chaos ($LLE>0.01$) drives 
the CLT Pdfs towards the standard Gaussian attractor, as shown in Fig.4.
Again we plot both $r(t)$ and $LLE(t)$ in the 
top panels, while in the middle row and in the lower row we plot, respectively, the Pdfs 
of the same run for different values of $\delta$ (200 and 3000) at $n=20$ and $n$ (220 and 1000) at $\delta=200$. 
In this case, increasing either $\delta$ or $n$, the Pdfs converge towards a Gaussian shape 
starting from the central part and slowly involving the tails. Again, a standard Gaussian
is also plotted for comparison.
It is important to stress that here the fully chaotic region fills almost all the partially synchronized regime and the 1st-order phase transition is characterized by a coexistence of the incoherent and synchronized phases which gives rise also (see full line in Fig.2) to the appearance of metastable states \cite{kura-hmf2} (in analogy with the QSS regime of the HMF model). In Fig.4 we avoid these metastable states and we show a typical run whose $LLE$ stationary value is consistent with the average value corresponding to the peak in Fig.1 (i.e. $LLE\sim0.06$ at $K=2.53$).

Finally, in Fig.5, we further increase the coupling up to $K=3$, shifting in the fully synchronized regime. Here the locked cluster tends to include almost all the oscillators, whose angles do not change in time any more (we remind that
the synchronized cluster tends to stop, being $\Omega\sim0$). In this situation one has $r>0.8$ while the $LLE$ is slightly negative, and the CLT Pdfs coincide (for large $\delta$ or $n$) with rescaled angles distributions (being $y_i\sim \theta_i$ for each $i=1,...,N$).
\begin{figure}  
\begin{center}
\epsfig{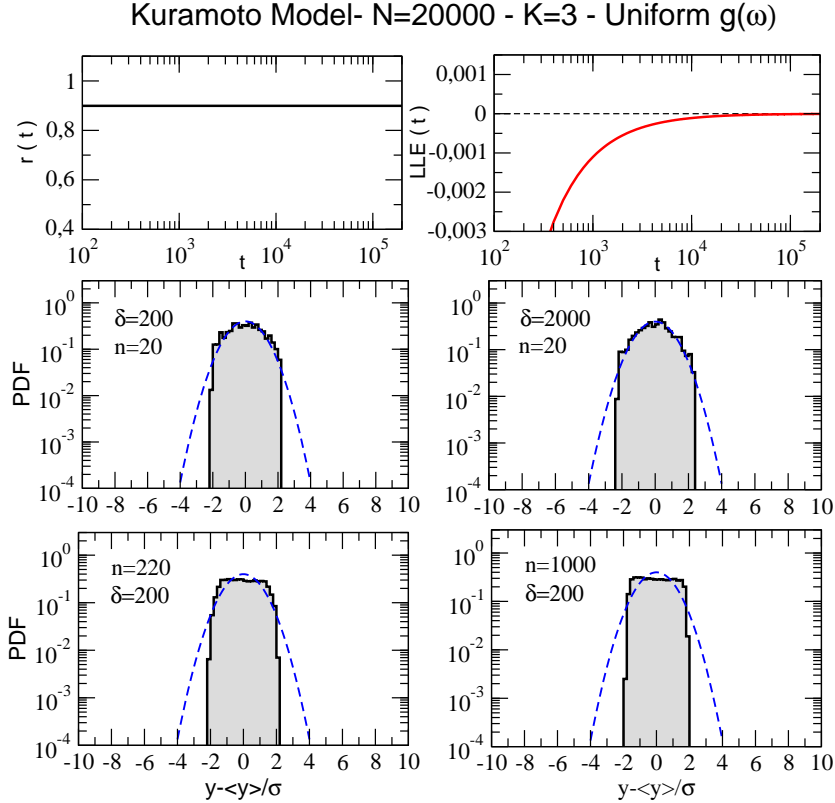}
\end{center}
\caption{Uniform $g(\omega)$, $N=20000$ and $K=3.0$ (in this case $K_C=2.54$). In the fully synchronized phase
the CLT Pdfs coincide with the angles distributions of the oscillators in the locked
cluster. Gaussian curves are also reported for comparison. See text. } 
\label{fig5}
\end{figure}
Summarizing, in the case of a uniform $g(\omega)$ distribution of the natural frequencies, 
the previous scenario confirms that only in the incoherent regime at the "edge of chaos", with vanishing $LLE$, robust $q$-Gaussian-like attractors appear in the CLT Pdfs of the Kuramoto model, then leaving room, when the level of chaos increases, to the standard Gaussian behavior predicted by the CLT.
Let us to explore, now, what happens for a Gaussian $g(\omega)$ distribution, when a 2nd-order phase transition occurs.

\begin{figure}  
\begin{center}
\epsfig{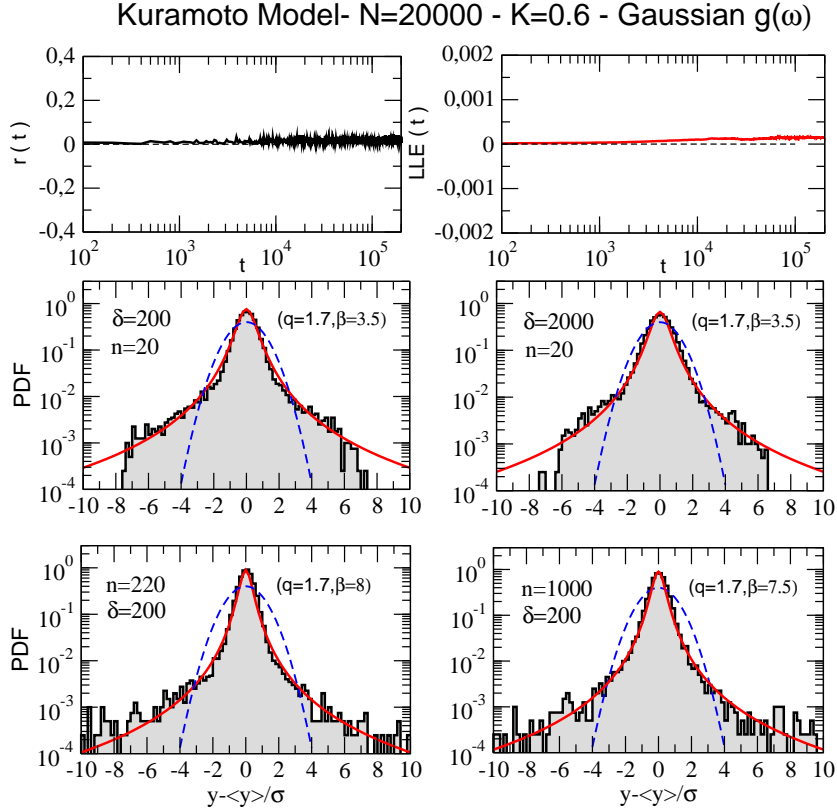}
\end{center}
\caption{Gaussian $g(\omega)$: CLT Pdfs for a Kuramoto system with $N=20000$ and $K=0.6$ (in this case $K_C=1.59$), 
are reported for increasing values of $\delta$ at fixed $n$ (middle row) and for increasing 
values of $n$ at fixed $\delta$ (lower row): as happened for the uniform case in Fig.3, 
being in the "edge of chaos" regime, robust fat tailed attractors appear again, 
well fitted by $q$-Gaussians (full lines) with $q=1.7$.
Standard Gaussians with unitary variance (dashed lines) are reported for comparison. 
In the upper row the order parameter (on the left) and the LLE (on the right) 
are plotted for comparison as function of time, again after a transient of 100 time-steps. 
See text. } 
\label{fig6}
\end{figure}

\subsection{Gaussian $g(\omega)$ distribution}

Considering a $\sigma=1$ Gaussian $g(\omega)$ distribution of the natural frequencies,
the region of interest in the diagram of Fig.2 shifts towards the vertical section 
at $\sigma=1$ (corresponding to a Gaussian-like distribution, but quite similar to the 
true Gaussian case). Notice that - at variance with what happened along the $\sigma=5$ section - 
the partially synchronized regime is here mostly situated above the critical line,
a footprint of the second order phase transition. 
As in the previous section we start studying the region of weak chaos in the incoherent phase.

In the upper row of Fig.6 we plot the time behavior of the order parameter and the
$LLE$ for a typical run at $K=0.6$, and one can see that both vanish as expected 
(in particular $LLE \sim 10^{-4}$).     
In the other panels CLT Pdfs are reported for increasing values of $\delta$ (200 and 2000) 
at fixed $n$=20 (middle row) and for increasing values of $n$ (220 and 1000) at fixed 
$\delta$=200 (lower row).
As in Fig.3, we found again robust fat tailed Pdfs that in this case can be fitted again very well 
by $q$-Gaussian curves (full lines) with $q=1.7$ and slightly different values of $\beta$. 
The usual standard Gaussian with unitary variance (dashed line) is reported for comparison.
\begin{figure}  
\begin{center}
\epsfig{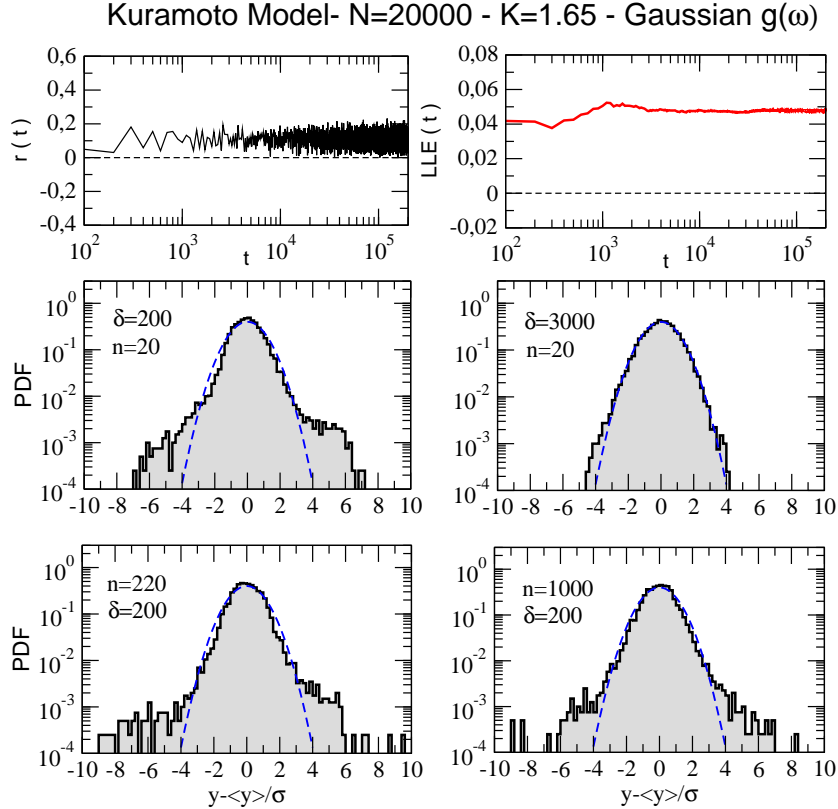}
\end{center}
\caption{Gaussian $g(\omega)$, $N=20000$ and $K=1.65$. The fully chaotic regime produces 
Gaussian-like Pdfs increasing either $\delta$ or $n$. Gaussian curves are
also reported for comparison. See text. } 
\label{fig7}
\end{figure}

In the simulations showed in Fig.7 we increase the coupling in the fully chaotic regime, 
up to the value $K=1.65$, very near to critical point, where the mixing effects indicated 
by the peak in the $LLE$ (see Fig.1, open circles line) slowly destroy the correlations 
and drive the CLT Pdfs towards a Gaussian-like shape which again starts from the central part, 
as in the uniform $g(\omega)$ case. 
Such a behavior is evident increasing either $\delta$ (at $n=20$) or $n$ 
(at $\delta=200$). In both cases we plot for comparison the usual Gaussian curve.
In this situation no coexistence of different phases is observed, even if metastable states are still present
for $1.65<K<1.75$. Therefore, again, we avoid these states and choose a typical run whose $LLE$ stationary value is consistent with the corresponding peak in Fig.1 ($LLE\sim0.05$ at $K=1.65$).

\begin{figure}  
\begin{center}
\epsfig{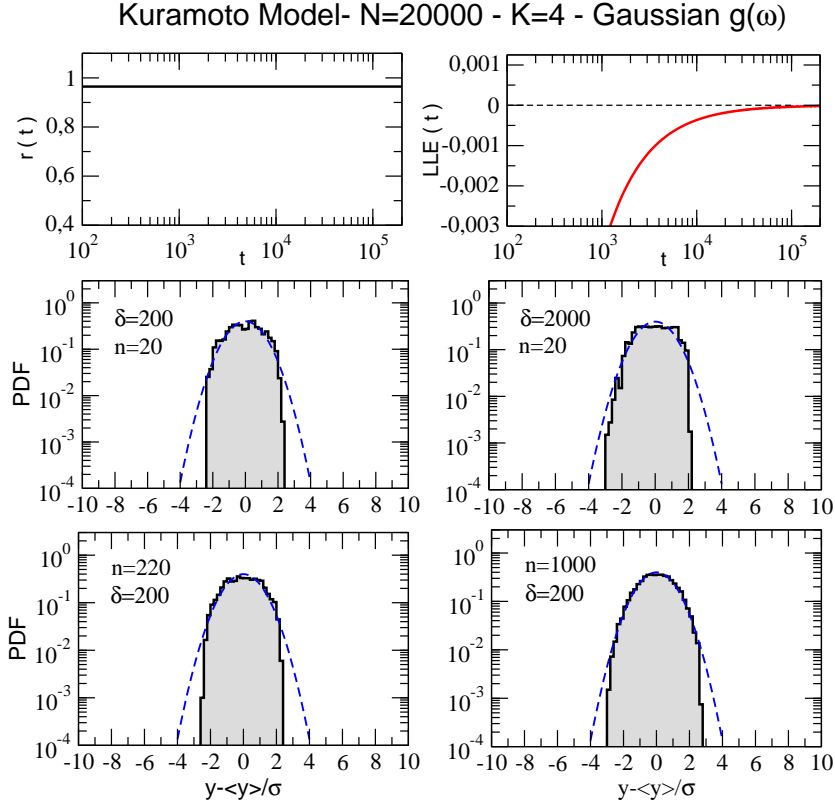}
\end{center}
\caption{Gaussian $g(\omega)$, $N=20000$ and $K=4.0$. In the fully synchronized phase
the CLT Pdfs end up to coincide with the angles distributions of the oscillators in the locked
synchronization cluster. Gaussian curves are also reported for comparison. See text. } 
\label{fig8}
\end{figure}

Finally, in Fig.8 we further increase $K$ in the fully synchronized regime up to $K=4$ and 
we observe a behavior analogous to that of Fig.5: being $r>0.8$ and $LLE\sim0$ but slightly negative, 
the CLT Pdfs tend to coincide with the angles distributions, assuming here a slightly asymmetric Gaussian-like 
shape for large values of $\delta$ and $n$.
Summarizing, also in the case of a Gaussian $g(\omega)$ distribution of the natural frequencies, 
only the vanishing Lyapunov exponent regime characterizing the incoherent phase gives rise to 
robust $q$-Gaussian-like attractors in the CLT Pdfs. 

\section{Conclusions}

We have studied the relationship between CLT and chaos in the Kuramoto model.
By investigating, along the deterministic  time evolution, the behavior of the sums  of the phases of the oscillators, we have  shown that when chaos is fully developed the CLT is satisfied and Gaussian Pdfs emerge as  attractors. This is the expected result  predicted  by the standard CLT  and obtained in other  Hamiltonian models  recently explored \cite{CLT-hmf1,CLT-hmf2}, but also in dissipative systems like the logistic map \cite{CLT-maps1,CLT-maps2}.
On the other hand, when the system is at the "edge of chaos", i.e.  well  below the  critical value of the coupling, where the LLE tends to zero, the standard CLT no longer applies.  As observed  in the recent literature for other models, in this case  the Pdfs  have very fat tails  and the curves  tend
to attractors  which can be very well fitted with q-Gaussians. Although we have presented only numerical simulations which cannot be considered a rigorous proof, the results are in line with the recently generalized version of the CLT proposed by Tsallis \cite{tsallis0} and provide further  support to the claim that q-Gaussian-like attractors  appear very often in situations  where one has vanishing Lyapunov exponents so that mixing and  ergodicity are violated.

\section{Acknowledgments}

We thank  Marcello Iacono Manno for the technical support to run  our codes on the TRIGRID platform.

\end{document}